\documentclass[prb, twocolumn,preprintnumbers ,amsmath, amssymb, superscriptaddress, aps]{revtex4-1}

\usepackage{color}
\usepackage{amsmath}
\usepackage{pifont}
\usepackage{amssymb}  
\usepackage{bbold}
\usepackage{float}
\usepackage{tikz}
\usepackage{makecell}
\usepackage{subfigure}
\usepackage{pifont}   
\usepackage{graphicx} 
\usepackage{dcolumn}  
\usepackage{bm}       
\usepackage{multirow} 
\usepackage{hyperref}
\usepackage{placeins}

\begin{document}

\renewcommand\floatpagefraction{0.8} 
\renewcommand\topfraction{0.8}       

\title{Group theory study of the vibrational modes and magnetic order in the topological antiferromagnet MnBi$_2$Te$_4$}

\author{Martin Rodriguez-Vega}
\affiliation{Department of Physics, The University of Texas at Austin, Austin, TX 78712, USA}
\affiliation{Department of Physics, Northeastern University, Boston, MA 02115, USA}

\author{A. Leonardo}
\affiliation{Donostia International Physics Center, San Sebastian, Spain}
\affiliation{Applied Physics Department II, University of the Basque Country UPV/EHU, Bilbao, Spain.}

\author{Gregory A. Fiete}
\affiliation{Department of Physics, Northeastern University, Boston, MA 02115, USA}
\affiliation{Department of Physics, Massachusetts Institute of Technology, Cambridge, MA 02139, USA}

\date{\today}

\begin{abstract}
We employ group theory to study the properties of the lattice vibrations and magnetic order in the antiferromagnetic topological insulator MnBi$_2$Te$_4$, both in bulk and few-layer form. In the paramagnetic phase we obtain the degeneracies, the selection rules, and real-space displacements for the lattice vibrational modes for different stacking configurations.  We discuss how magnetism influences these results.  As a representative example, we consider a double septuple layer system and obtain the form of the magnetic order allowed by the symmetries of the crystal.  We derive the allowed coupling terms that describe the paramagnetic to antiferromagnetic transition. Finally, we discuss the implications of our results for Raman scattering and other optical measurements. Our work sets the stage for a deeper understanding of the interplay of lattice modes, band topology, and magnetic order in MnBi$_2$Te$_4$ and other symmetry-related materials.%
\end{abstract}
\maketitle

\section{Introduction}

The discovery of the integer quantum Hall effect \cite{vonklitzing1980} and its subsequent explanation in terms of the underlying topology of the band structure \cite{thouless1982} has led to a new perspective on phases of matter, exemplified by topological insulators~\cite{thouless1982,kane2005,fu2007,hasan2010,Moore2010,qi2011} and certain unconventional superconductors~\cite{Kitaev_2009,Beenakker_2016,Sato_2017}. The quantum spin Hall effect in two dimensions~\cite{konig2007}, time-reversal invariant topological insulating phases in three dimensions~\cite{hsieh2008,Zhang2009,Xia2009}, topological crystalline insulators~\cite{fu2011}, higher-order topological phases with gapped bulk and surfaces but conducting hinges~\cite{Schindler_2018,Schindler_2018t,Benalcazar61}, and fractionalized variants of these phase\cite{Maciejko:np15,Stern:arcmp16,Kargarian:prl13,Ruegg:prl12,Maciejko:prl14,Maciejko:prl10} are all outgrowths of topological ideas rooted in the quantum Hall effects. The exploration of topological phases has not been limited to condensed matter systems: there have been advances in optical lattices~\cite{Goldman_2016}, photonics ~\cite{Lu2016}, mechanical and acoustic systems~\cite{Huber2016,Ma2019}, and out-of-equilibrium systems~\cite{oka2019,Rudner2020}.

The interplay of topological phases with correlated states of matter, such as magnetism, is currently one of the most interesting aspects in the field~\cite{Tokura2019}. Magnetic topological insulators (MTIs) can host several interesting phenomena such as the quantized anomalous Hall (QAH) effect, characterized by a robust quantized Hall conductivity in the absence of an applied magnetic field~\cite{haldane1988,Ohgushi2000}. In two-dimensions, the QAH effect was predicted~\cite{yu2010} and observed in thin-film magnetically doped topological insulators \cite{Chang2013,Yu61}. In three-dimensions, symmetry protection can lead to an axion insulating phase, where the axion term in the electromagnetic Lagrangian~\cite{xiao2008} is, $ {\cal L}_{\theta} =  \theta e^2 /(2 \pi h) \boldsymbol E \cdot \boldsymbol B$, with $\theta = \pi$ protected by the bulk inversion symmetry induces a quantized magnetoelectric effect (in units of the finite structure constant). These axion insulators have been realized in doped (Bi,Sb)$_2$Te$_3$ heterostructures, as revealed by electrical transport measurements~\cite{PhysRevLett.120.056801,Mogi2017,Chang2015}. 
\begin{figure}[t]
	\begin{center}
		\includegraphics[width=8.0cm]{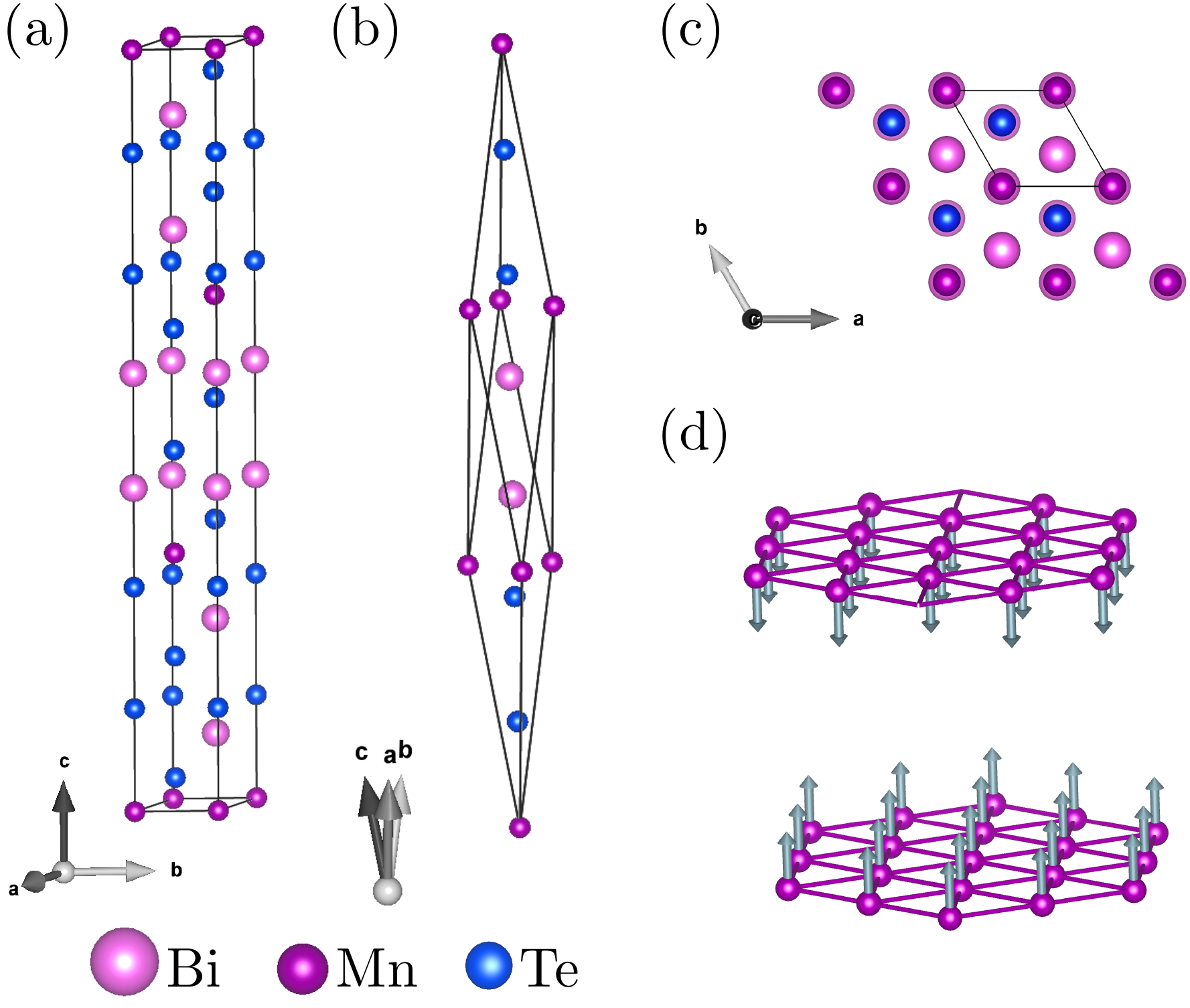}
		\caption{(Color online) Paramagnetic MBT with (a) hexagonal and (b) rombohedral unit cells, which contain a single Te-Bi-Te-Mn-Te-Bi-Te septuple layer. (c) Crystal structure viewed along the $c$-axis in the hexagonal setting. (d) Antiferromagnetic ordering of bulk MBT between two adjacent Mn layers, showing only the magnetic atoms for clarity. This figure was created with VESTA~\cite{vesta}.}
		\label{fig:param_unit_cell}
	\end{center}
\end{figure}

The search for compounds which can intrinsically host magnetic topological phases led to the recent theoretical prediction~\cite{Otrokov2019, Zhang_2019, Li_2019} and experimental realization~\cite{Otrokov2019, Gong_2019} of MnBi$_2$Te$_4$ (MBT) and related compounds MnBi$_4$Te$_7$~\cite{Hu_2020,doi:10.1116/1.5122702,ALIEV2019443}, MnBi$_6$Te$_{10}$~\cite{ALIEV2019443} and MnBi$_8$Te$_{13}$~\cite{Hueaba4275}. Bulk MBT is a van der Waals compound (Fig. \ref{fig:param_unit_cell} shows the crystal structure) which exhibits ferromagnetic (FM) order within each Mn plane and antiferromagnetic (AFM) order between the layers below $T_N \approx 24$K~\cite{Otrokov2019,Yan2019}. 

In its paramagnetic phase, MBT is a strong topological insulator protected by time-reversal symmetry with a gapped bulk and protected surface states~\cite{Chen2019_param}. Below the critical temperature, AFM order sets in and  MBT transitions into an AFM TI phase protected by the composition of time-reversal and a translation in the $c$-axis by half a lattice constant~\cite{Otrokov2019}. Theoretically, it was proposed that when the sample is terminated along the (111) surface, the symmetry is broken at the surface and gapped surface states are obtained~\cite{PhysRevB.81.245209,Otrokov2019, Zhang_2019}. Subsequent angle resolved photon emission spectroscopy (ARPES) measurements indeed observed the predicted gapped bulk and surface states~\cite{PhysRevB.100.121104,PhysRevX.9.041040,PhysRevResearch.1.012011,Otrokov2019}.  However, the relation of the surface state gap to the magnetic order is still under investigation~\cite{PhysRevB.100.121104,PhysRevX.9.041040,PhysRevResearch.1.012011}, including reports of gapless surface states using high-resolution ARPES~\cite{PhysRevX.9.041038,PhysRevB.101.161109}. 

The topological nature of the system has been revealed through magneto-transport measurements which have shown the QAH effect in five septuple-layer MBT at low temperatures~\cite{Deng_2020} and axion insulating states at zero magnetic field in six septuple-layer samples~\cite{Liu2020}. Finally, as a function of an applied magnetic field parallel to the crystal $c$-direction (hexagonal cell), bulk MBT presents a spin-flop transition at 3.5 T~\cite{Otrokov2019,Deng_2020}, which characterizes the strength of the interlayer interaction.

Given that MBT is a van der Waals compound~\cite{Otrokov2019}, it is possible to exfoliate septuple layers (SL) and form few-layer stable structures~\cite{Deng_2020,Liu2020}. As shown by experiments ~\cite{Deng_2020,Liu2020} and first-principles calculations~\cite{Otrokov2019_thickness}, the topological and magnetic properties of MBT depend on the number of layers~\cite{Otrokov2019_thickness}. These thickness-dependent features open up the possibility to combine MBT with other van der Waals materials to create heterostructures with designer properties~\cite{Fueaaz0948}. 

In parallel to these developments on topology and magnetism, there is increasing interest in controlling magnetic and other correlated states using light~\cite{oka2009, lindner2011,FORST201324,subedi2014, Mentink2015, gu2017, juraschek2017,Juraschek2017b,subedi2017,Babadi2017,liu2018,Juraschek2018,Juraschek2019,Juraschek2020,juraschek2019phonomagnetic,kalsha2018,gu2018,fechner2016,Hejazi2019,Sentef2016,Sentef2017,Tancogne2018,chaudhary2019phononinduced,Chaudhary2019,vogl2020effective, vogl2020floquet, PhysRevX.9.021037,baldini2020,PhysRevLett.107.216601,PhysRevLett.116.016802,nguyen2020photoinduced,asmar2020floquet}. For this application, low-frequency (THz regime) drives are desirable, since they minimizes heating issues inherent in driven interacting systems. In this regime, the phonons become the relevant degrees of freedom and their non-linear interactions provide a mechanism to manipulate the lattice in a controlled manner. The induced lattice distortions can in turn influence a correlated state, such as magnetism, and induce transitions on demand~\cite{Fausti189,Mankowsky2014,mitrano2016,rodriguezvega2020phononmediated,juraschek2019phonomagnetic,kalsha2018,gu2018,lingos2020lightwave}. 

Motivated by the dependence of the topological properties of MBT on the number of layers--as shown in experiments--and its potential application in light-driven heterostructures, in this work we use group theory to analyze the properties of the lattice vibrations as a function of the number of layers and the symmetry aspects of the magnetic transitions. 

Group theory approaches have been used to study two-dimensional transition metal dichalcogenides~\cite{Ribeiro_Soares_2014} and phosphorene multilayer systems~\cite{Ribeiro_Soares_2015}. Here, we consider both bulk and few-septuple layer MBT in its paramagnetic phase assuming different stacking configurations and the symmetry allowed magnetic phases. Our results could be used to interpret Raman measurements and distinguish stacking patterns though the angular dependence of the modes. 

Our work is organized as follows. In Sec. \ref{sec:latt}, we study the bulk vibrational modes in the paramagnetic phase. For double-SL MBT (d-MBT), we consider  several stacking configurations between the SLs, obtain the space group, the phonon selection rules, and lattice displacements.  For tri-SL MBT (t-MBT), we consider the bulk-like ABC stacking and selected misaligned structures and determine their impact on the phonon properties. In Sec. \ref{sec:magneticmbt}, we study the symmetry aspects of the transition from the paramagnetic to the AFM phase focusing on d-MBT. Finally, in Sec. \ref{sec:conclusions}, we discuss the implications of our analysis to experimental measurements and present our conclusions. Throughout this work, we employ the group theory packages \small{ISOTROPY}~\cite{isotropy}, GTPack~\cite{gtpack1,gtpack2} and the Bilbao crystallographic server~\cite{genpos}.

\section{Vibrational modes in the paramagnetic phase}
\label{sec:latt}

\subsection{Bulk}

In this section, we consider  MBT in the paramagnetic phase, where the crystal symmetries fully determine the properties of the vibrational modes. Figure \ref{fig:param_unit_cell} shows the conventional unit cell (the primitive unit cell contains one Mn, two Bi, and four Te atoms, for a total of $ N = 7 $ atoms.). The crystal is formed by  Te-Bi-Te-Mn-Te-Bi-Te septuple layers (SLs) staked on top of each other. Bulk MBT belongs to the space group $R \bar 3 m$ $(166)$~\cite{C3CE40643A, Otrokov2019} with point group $D_{3d}$. 

The character table for $D_{3d}$ is shown in Appendix~\ref{app:tables}.  The symmetry operations of the point group are $\{E,2C_3,3C'_{2},i,S_6, 3\sigma_d\}$, and the  irreducible representations (irreps) are $A_{1g}, A_{2g}, E_g,A_{1u}, A_{2u}, E_u$. Notice that inversion is a symmetry of the lattice and therefore the irreps can be labeled by their transformation properties under inversion: $g$ for even and $u$ for odd transformation. 

\begin{figure}
	\begin{center}
		\includegraphics[width=8.5cm]{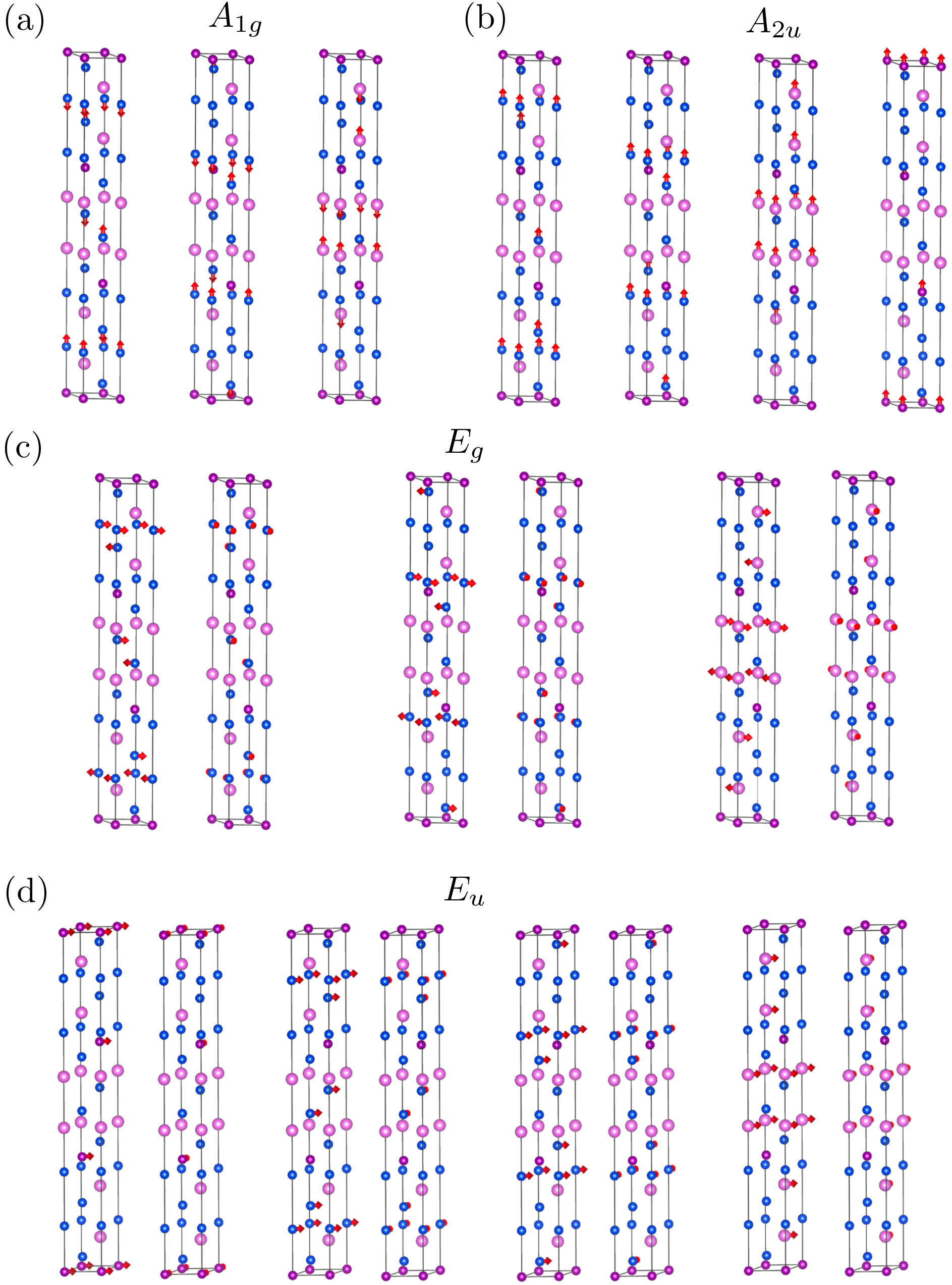}
		\caption{(Color online) Lattice vibrational modes for bulk MBT in the paramagnetic phase. The red arrow indicates the direction of the atomic motion. The color coding of the atoms is the same as in Fig.\ref{fig:param_unit_cell}. This figure was created with VESTA~\cite{vesta}. }
		\label{fig:latt_vib}
	\end{center}
\end{figure}

The irreducible representations of the vibrational modes are given by $\Gamma_{vib.}  = \Gamma^{equiv} \otimes \Gamma_{vec}$~\cite{GroupTheoryDress2008}, where $\Gamma^{equiv}$ is the equivalence representation, determined by the atoms that do not change positions upon application of the group symmetry operations. Here $\Gamma_{vec} = E_u \oplus A_{2u}$ corresponds to the representation of the vector in the point group $D_{3d}$. Then, we find $\Gamma^{equiv}= 4 A_{1g} \oplus 3A_{2u}$, which leads to  the lattice vibration representation
\begin{equation}
\Gamma_{vib.}  = 3 A_{1g} \oplus 4 A_{2u} \oplus 3E_g \oplus 4E_u,
\end{equation}
upon expansion into irreps. We have $7$ non-degenerate ($A$) and $7$ doubly-degenerate ($E$) phonon modes for a total for $3N = 21$ modes. Since the point group contains inversion, we can distinguish between infrared (IR)  and Raman active modes. The infrared modes transform as a vector $\Gamma_{vec}$,  while Raman modes transform as $\Gamma_{vec} \otimes \Gamma_{vec} = 2 A_{1g} \oplus A_{2g} \oplus 3 E_g$. In this case, $9$ modes are Raman active (3 with totally symmetric $A_{1g}$ representation and 3 with doubly-degenerate $E_g$ representation) and $12$  IR active modes, including the acoustic modes  ($A_{2u} \oplus E_u$)~\cite{kroumova2003}.

In a Raman scattering experiment, the contribution of the Raman mode $\nu$ to the total intensity of the signal is $I_{\nu}\propto \left|\sum_{\alpha, \beta= \{x,y,z\}} e_{\mathrm{in}}^{\alpha} R_{\alpha \beta}^{\nu} e_{\mathrm{out}}^{\beta}\right|^{2} $, where  $e_{\mathrm{in}/\mathrm{out}}^{\alpha}$  are the polarization vectors of the incoming and outgoing light and $R_{\alpha \beta}^{\nu} $ are the components of the Raman tensor for mode $\nu$. The total intensity has the form
$I(\omega)=I_{0} \sum_{\nu} g_{\nu} I_{\nu} \delta\left(\omega-\omega_{\nu}\right)$, where  $\omega$ is the probe frequency, and $g_{\nu} $ is an occupation pre-factor which depends on whether we have a Stokes or an antiStokes process. $I(\omega)$ can be derived from third-order perturbation theory, for detailed derivations see Ref.~\onlinecite{yu2010fundamentals}.  

Therefore, the selection rules for the Raman modes are determined by the Raman tensors $R$, since the Raman intensity can be non-zero in a given Raman setup is the corresponding elements of the Raman tensor are non-zero. For paramagnetic MBT, the Raman tensors $R$ are given by~\cite{loudon1964}

\begin{align}
R(A_{1g}) & = \begin{pmatrix}
a & 0 & 0\\
0 & a & 0\\
0 & 0 & b
\end{pmatrix}\\
R^{(a)}(E_{g}) & = \begin{pmatrix}
c & 0 & 0\\
0 & -c & d\\
0 &  d & 0
\end{pmatrix}, R^{(b)}(E_{g}) & = \begin{pmatrix}
0 & -c & -d\\
-c  & 0 & 0\\
-d & 0 & 0
\end{pmatrix}.  
\end{align}

The real-space displacements, which bring the dynamical matrix into a block-diagonal form where each block corresponds to a different irrep~\cite{GroupTheoryDress2008, gtpack2}, can be obtained constructing  the projection operators~\cite{GroupTheoryDress2008, gtpack2} 
\begin{equation}
\hat P^{(\Gamma_n)}_{kl} = \frac{l_n}{h} \sum_{C_\alpha} \left( D_{kl}^{(\Gamma_n)}(C_\alpha) \right)^* \hat P(C_\alpha),
\end{equation}
and applying them to arbitrary displacements. Here, $\Gamma_n$ are the  irreps, $C_\alpha$ are the elements of the group, $D_{kl}^{(\Gamma_n)}(C_\alpha) $ is the irreducible matrix representation of element $C_\alpha$  (which are obtained with GTPack~\cite{gtpack1,gtpack2}), $h=12$ is the order of $D_{3d}$, and $l_n$ is the dimension of the irreducible representation. Finally, $\hat P(C_\alpha)$ are $3N\times 3N$ matrices that form the displacement representation. Here, we employ the suite \small{ISODISTORT} to efficiently find the real space displacements~\cite{isotropy,isodistord} and Fig. \ref{fig:latt_vib} shows the $21$ modes along with their irrep label. Interestingly, for $A_{1g}$ and $E_{g}$ modes, the magnetic Mn atoms do not participate in the lattice vibrations. 

In summary, in this subsection we characterized the vibrational modes of bulk paramagnetic MBT. We obtained the Raman and IR active modes, their selection rules determined by the Raman tensors, and found the real-space vibrations which bring the dynamical matrix to block-diagonal form. In the next subsection, we consider few-layer MBT, and discuss how these properties vary as a function of the number of layers and different stacking configurations.

\subsection{Few-septuple layers MBT}

The weak van der Waals nature of the forces holding the layers of MBT together allows one to obtain samples down to the single SL limit (Te-Bi-Te-Mn-Te-Bi-Te)~\cite{Deng_2020,Liu2020}. In this section, we analyze the properties of the vibrational modes for the case of a single SL (m-MBT), double SL layer (d-MBT), and triple SL systems (t-MBT). 

\begin{figure}
	\begin{center}
		\includegraphics[width=8.5cm]{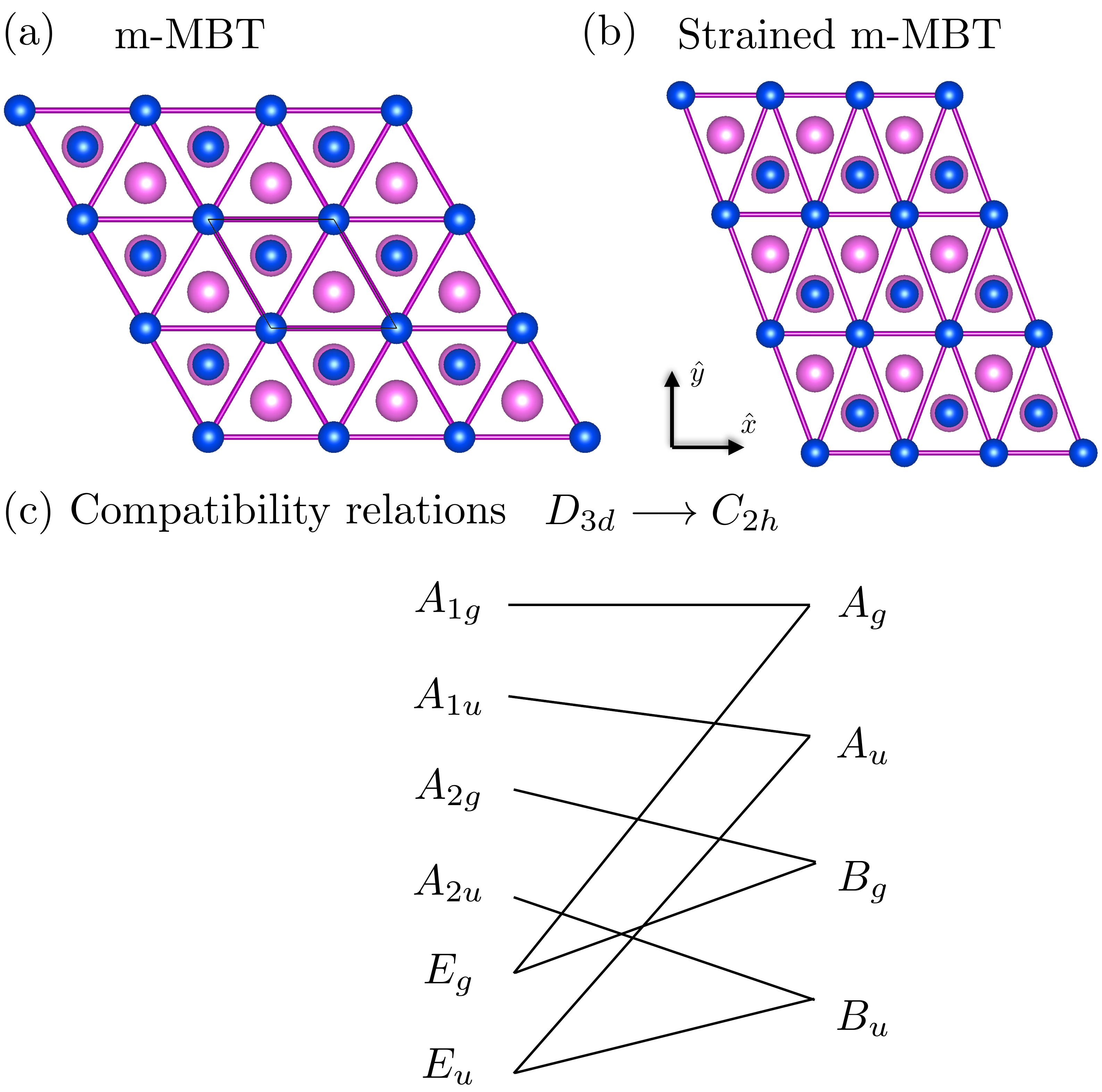}
		\caption{(Color online) Crystal structure for (a) pristine m-MBT and (b) strained m-MBT. The color coding of the atoms is the same as in Fig.\ref{fig:param_unit_cell}. These figures were created with VESTA~\cite{vesta}.  (c) Compatibility relations for the group-subgroup pair.}
		\label{fig:mono_strain}
	\end{center}
\end{figure}

\textit{Single SL}. For m-MBT the space group reduces to $P\bar 3m1$ ($164$), which can be understood from the lattice structure since m-MBT loses the symmetries composed with translations in the $c$-direction (hexagonal axis). The crystal structure is shown in Fig. \ref{fig:mono_strain}(a), viewed along the $c$-axis (parallel to the cartesian $z$ axis). Nonetheless, the point group is $D_{3d}$ as in bulk. Te atoms are located at Wyckoff's positions 2d $ (1/3,2/3,z)$ and 2c $(0,0,z)$, Bi atoms at 2d $(1/3,2/3,z)$, and the Mn atom at 1b $(0,0,1/2)$. Since m-MBT has the same number atoms in the primitive unit cell as bulk MBT, the representation for the lattice vibration is also $\Gamma_{vib.} = 3 A_{1g} \oplus 4 A_{2u} \oplus 3E_g \oplus 4E_u$ . 

Now, let us consider the effect of in-plane strain on m-MBT, which could result from placing the monolayer on a substrate. We assume that strain is applied in the cartesian $y$-direction, resulting in the distorted structure shown in Fig. \ref{fig:mono_strain}(b). This distortion leads to a lowering of the symmetry to the space group $C2/m$, with point group $C_{2h}$. The compatibility relations are shown in Fig. \ref{fig:mono_strain}(c), which relate irreps in the $D_{3d}$ and $C_{2h}$ point groups. The representation for the lattice vibration becomes $\Gamma_{vib.} = 6 A_{g}+4 A_{u}+3 B_{g}+8 B_{u}$, with Raman tensors~\cite{loudon1964}

\begin{align}
R(A_{g}) & = \begin{pmatrix}
a & d & 0\\
d & b & 0\\
0 & 0 & c
\end{pmatrix},
R(B_{g})  = \begin{pmatrix}
0 & 0 & e\\
0 & 0 & f\\
e &  f & 0
\end{pmatrix}.  
\label{eq:c2h_Raman}
\end{align}

\begin{figure}
	\begin{center}
		\includegraphics[width=7.0cm]{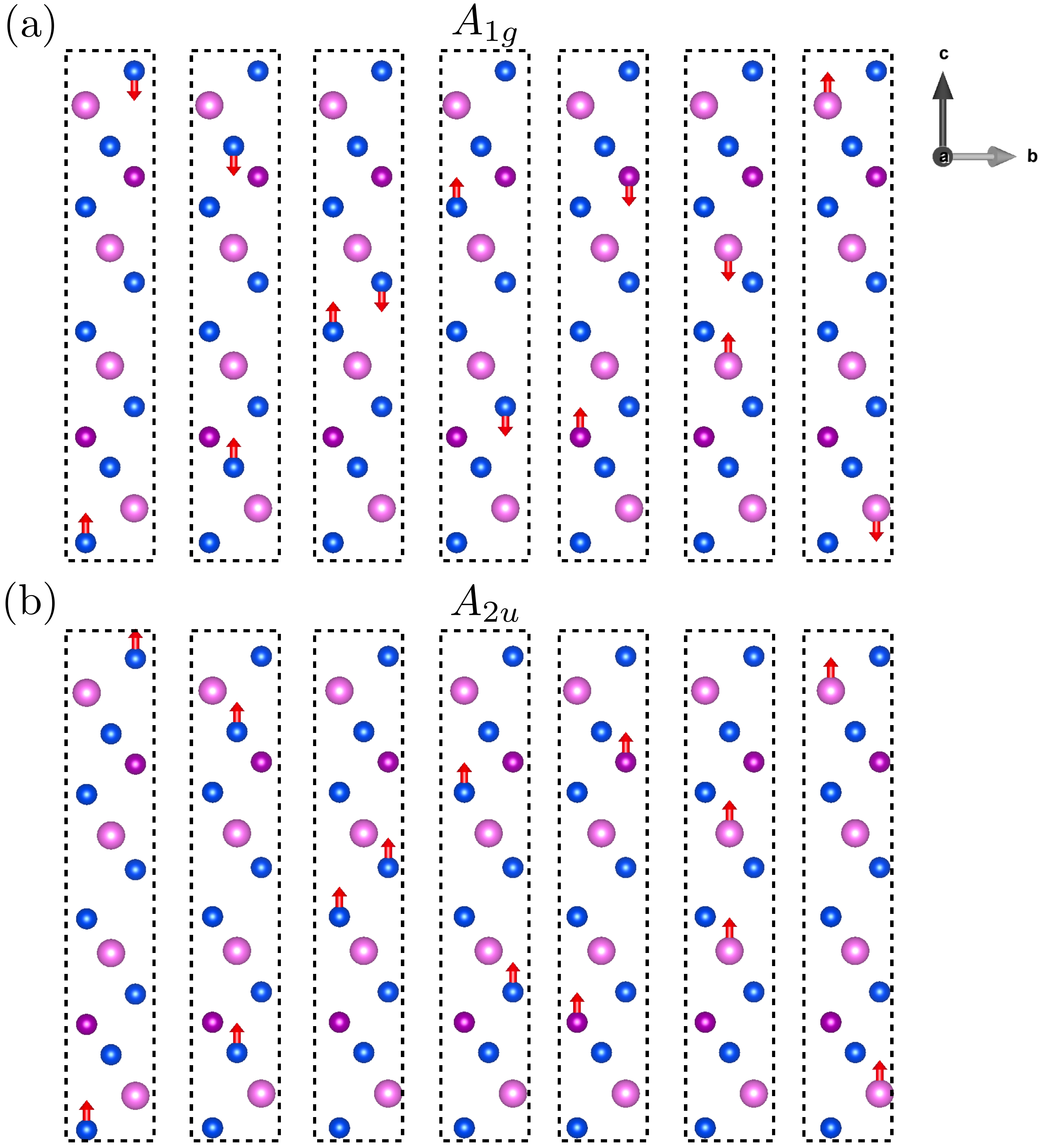}
		\caption{(Color online) Lattice vibrational modes with irreps $A_{1g}$ and $A_{2u}$ for d-SL MBT in the paramagnetic phase. The dashed boxes enclose the different modes. The color coding of the atoms is the same as in Fig.\ref{fig:param_unit_cell}.  This figure was created with VESTA~\cite{vesta}. }
		\label{fig:latt_vib_bi_a}
	\end{center}
\end{figure}

\textit{Double SL}. Now we consider the case of two SLs stacked on top of each other. First, let us consider d-MBT with AB stacking (same stacking pattern as in bulk), which shares the same space group with m-MBT, $P\bar 3m1$ ($164$) with point group $D_{3d}$ at wave vector $\Gamma$. In this case, the unit cell doubles in size, hosting now $N=14$ atoms in the unit cell. The Te atoms are located at $2d$; $(1/3,2/3,z)$ and $2c$; $(0,0,z)$ Wyckoff's positions, Mn atoms are placed at $2d$ positions, and Bi are at $2c$ and $2d$ positions with different $z$ coordinates. We find that the equivalence representation is given by $\Gamma^{equiv} = 7 A_{1g} \oplus 7 A_{2u}$, which leads to the lattice vibration representation 
\begin{equation}
\Gamma_{vib.}  = 7 A_{1g} \oplus 7 A_{2u} \oplus 7 E_g \oplus 7 E_u.
\end{equation}
We have $7 A_{1g}$ one-dimensional and $7 E_g$ doubly-degenerate Raman active modes. On the other hand, we have $6 A_{2u}$ one-dimensional and $6 E_u$ IR modes excluding the acoustic modes. The real-space lattice displacements that block-diagonalize the dynamical matrix with irreps $A_{1g}$ and $A_{2u}$ are shown in Fig \ref{fig:latt_vib_bi_a}. Contrary to the bulk case, Mn atoms do participate in the lattice displacements with  $A_{1g}$ and $E_{g}$  irreps ($E_{g}$ and $E_{u}$ are shown in Appendix (\ref{app:biSL})). Notice that modes where the layers are uniformly displaced away and towards each other are allowed. In particular, shear modes involving relative SL displacements within the plane posses $E_g$ representation. These types of modes have been used to theoretically manipulate the magnetic order in other bilayer van der Waals systems~\cite{rodriguezvega2020phononmediated} and induce ferroelectric switching in bilayer transition metal dichalcogenides~\cite{Park2019} via non-linear phonon processes.

\begin{table}[]
\begin{tabular}{|c|c|c|c|c|c|c|}
\hline
$AA$ stacking & $0$          & $1/4$      & $1/3$      & $1/2$      & $2/3$        & $3/4$      \\ \hline
$0$           & $P \bar 3m1$ & $P\bar 1$  & $P\bar1$   & $C2/m$     & $P\bar1$     & $P \bar 1$ \\ \hline
$1/4$         &              & $P \bar 1$ & $P \bar 1$ & $C2/m$     & $P \bar 1$   & $C2/m$     \\ \hline
$1/3$         &              &            & $P\bar1$   & $P \bar 1$ & $P \bar 3m1$ & $P \bar 1$ \\ \hline
$1/2$         &              &            &            & $C2/m$     & $P \bar 1$   & $C2/m$     \\ \hline
$2/3$         &              &            &            &            & $P\bar1$     & $P \bar 1$ \\ \hline
$3/4$         &              &            &            &            &              & $P\bar1$   \\ \hline
\end{tabular}
\caption{Space group for d-MBT with layers shifted by a fraction of the lattice vectors, starting from an AA stacking configuration. The table is symmetric, therefore we omit the transpose slots for clarity.}
\label{tab:bmbt-sgps}
\end{table}

Because the fabrication processes for d-MBT can lead to different stacking patterns when placing one SL on top of the other, we now consider deviation from the bulk ABC pattern and its consequence on the symmetry group. In Table \ref{tab:bmbt-sgps} we show the space group for different relative shifts between the layers. We construct this table by defining the shifted bilayer systems and evaluating the space group with \small{ISOTROPY}~\cite{isotropy, fyndsim}. For example, d-MBT with AA stacking, where the Mn atoms on opposite layers lie on top of each other belongs to the space group  $P\bar 3m1$ ($164$), same as with AB stacking. For most of the shifts considered, the symmetry reduces to $P\bar 1$, which only contains identity $\{ \mathbb  1 | (0,0,0)\}$ and inversion $\{ \mathcal I | (0,0,0)\}$ symmetries.  A set of shits leads to the space group $C2/m$, with symmetry operations:  $\{ \mathbb  1 | (0,0,0)\}$, reflection combined with  translation in the $z$-direction $\{ \sigma_{y} | (0,0,1/2)\}$, a pure translation $\{ \mathbb  1 | (1/2, 1/2,0)\}$, and reflection $\{ \sigma_{y} | (1/2,1/2,1/2)\}$. The special shift $(1/3,2/3)$ leads to the original ABC stacking with space group $P \bar 3m1$. Notice that all the shifts considered preserve inversion symmetry, and a generic shift can reduce the symmetry to $P1$. 

As a representative example of the systems with space group $P\bar 1$, we consider d-MBT-$(0,1/4)$. The point group at the $\Gamma$ point is $C_{i}$ which has two irreps: $A_g$ and $A_u$. The representation of the lattice vibrations is given by $\Gamma_{vib.} = 21 A_g \oplus 21A_u$, which corresponds to 21 Raman active modes, and 21 IR modes, including the 3 acoustic modes. Due to its low symmetry, this group does not impose restriction on the elements of the Raman tensor for $A_g$. d-MBT-$(0,1/2)$ corresponds to the space group $C2/m$ with point group $C_{2h}$ which has four one-dimensional irreps: $A_g$, $B_g$, $A_u$, and $B_u$. We find $\Gamma_{vib.} = 14 {A}_{{g}}\oplus7 {A}_{{u}}\oplus7 {B}_{{g}}\oplus14 {B}_{{u}}$. The Raman tensors for $C_{2h}$ are shown in Eq. \eqref{eq:c2h_Raman}. The different irreps for these configurations would lead to different behavior of the Raman spectrum as a function of the angle of the incident light. This could help to distinguish the stacking patterns in d-MBT. 

\textit{Triple and thicker SLs.} t-MBT, quadruple SL MBT, and quintuple SL MBT with ABC stacking share the same space group and point group as d-MBT. The successive increase in the number of atoms in the unit cell leads to an increase in the number of phonon modes. In Table \ref{tab:irreps_N}, we display the representation of the lattice vibrations $\Gamma_{vib.}$  for these few-layered systems.  

\begin{table}[]
\begin{tabular}{|c|c|c|}
\hline
Number of SLs & Space group  & $\Gamma_{vib.}$                                          \\ \hline
Bulk          & $R \bar 3 m$ & $3 A_{1g} \oplus 4 A_{2u} \oplus 3E_g \oplus 4E_u$       \\ \hline
1             & $P\bar 3m1$  & $3 A_{1g} \oplus 4 A_{2u} \oplus 3E_g \oplus 4E_u$       \\ \hline
2             & $P\bar 3m1$  & $ 7 A_{1g} \oplus 7 A_{2u} \oplus 7 E_g \oplus 7 E_u$    \\ \hline
3             & $P\bar 3m1$  & $10 A_{1g} \oplus 11 A_{2u} \oplus 10 E_g \oplus 11 E_u$ \\ \hline
4             & $P\bar 3m1$  & $14 A_{1g} \oplus 14 A_{2u} \oplus 14 E_g \oplus 14 E_u$ \\ \hline
5             & $P\bar 3m1$  & $17	 A_{1g} \oplus 18 A_{2u} \oplus 17 E_g \oplus 18 E_u$                                                          \\ \hline
\end{tabular}
\caption{Lattice vibration representation $\Gamma_{vib.}$ for few-SLs MBT with the bulk ABC stacking.}
\label{tab:irreps_N}
\end{table}

 \begin{table*}
\begin{tabular}{|c|c|c|c|c|c|c|c|c|c|c|c|c|}
\hline
$P \bar 3 m 11'$ & $\mathbb 1$ & $2 C_{3}$ & $3C_2$ & $\mathcal I$ & $2 \mathcal I C_3$ & $3 \mathcal I C_{2} $ & $\mathcal T$ & $2\mathcal T C_{3}$ & $3 \mathcal T C_2$ & $\mathcal T \mathcal I$ & $2 \mathcal T \mathcal I C_3$ & $3 \mathcal T \mathcal I C_{2}$ \\ \hline
$A_{1g}$    ($D_{3d}$)     & 1           & 1         & 1      & 1            & 1                  & 1                     & -1           & -1                  & -1                 & -1                      & -1                            & -1                              \\ \hline
$A_{1u}$     ($D_3$)    & 1           & 1         & 1      & -1           & -1                 & -1                    & -1            & -1                   &-1                  & 1                      & 1                            & 1                              \\ \hline
$A_{2g}$    ($C_{3i}$)     & 1           & 1         & -1     & 1            & 1                  & -1                    & -1           & -1                  & 1                  & -1                      & -1                            & 1                               \\ \hline
$A_{2u}$   ($C_{3v}$)      & 1           & 1         & -1     & -1           & -1                 & 1                     & -1           & -1                  & 1                  & 1                       & 1                             & -1                              \\ \hline
$E_{g}$      ($C_i$)    & 2           & -1        & 0      & 2            & -1                 & 0                     & -2           & 1                   & 0                  & -2                      & 1                             & 0                               \\ \hline
$E_{u}$     ($C_1$)     & 2           & -1        & 0      & -2           & 1                  & 0                     & -2           & 1                   & 0                  & 2                       & -1                            & 0                               \\ \hline
\end{tabular}
\caption{Irreducible correpresentations of the magnetic $P \bar 3 m 11'$ at the $\Gamma$ point. In parenthesis we show the point group associated with the crystallographic irreps. Only the odd co-representations are considered.}
\label{tab:IC}
\end{table*}

\begin{figure}[t]
	\begin{center}
		\includegraphics[width=8.5cm]{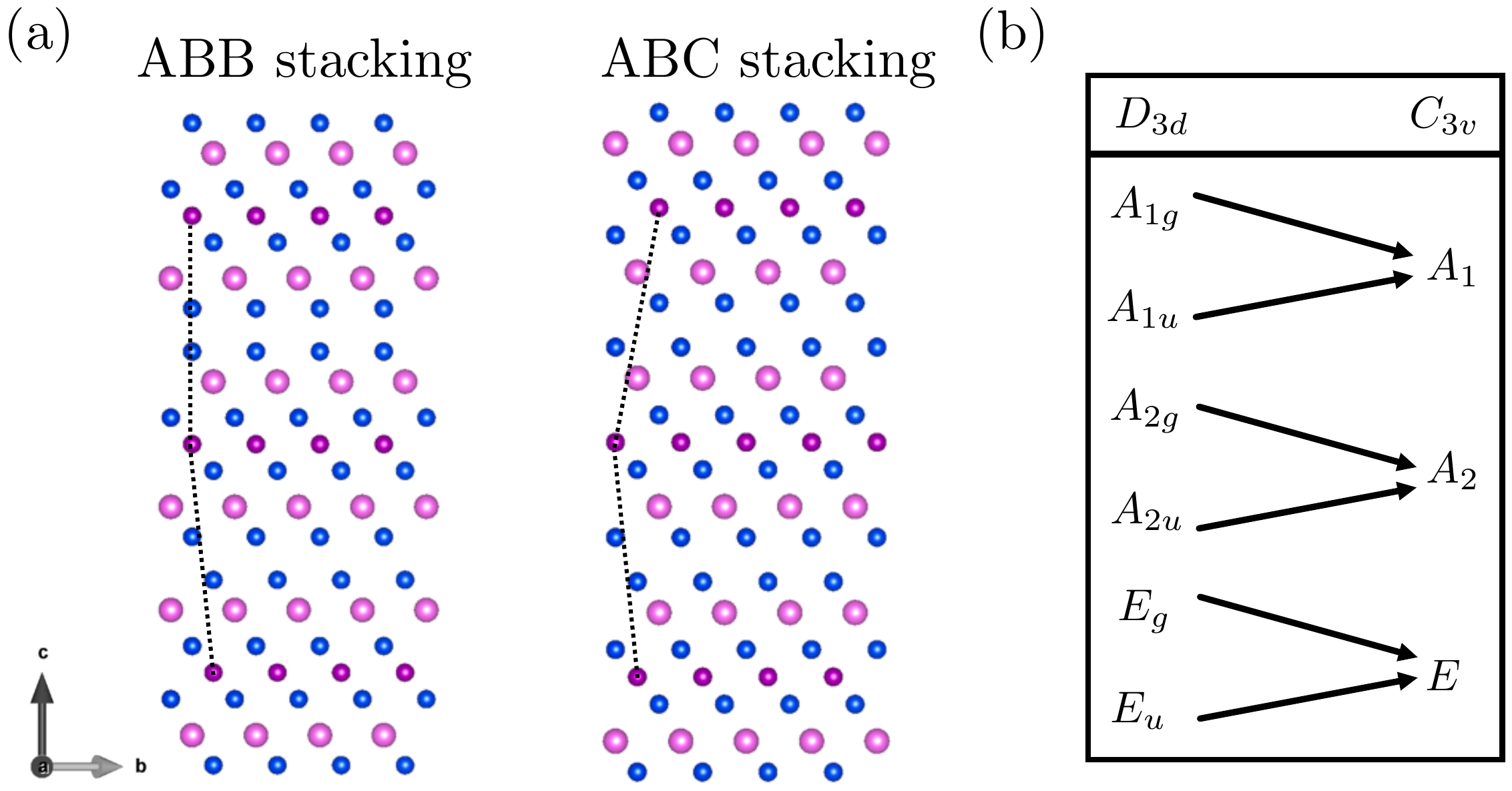}
		\caption{(Color online.) (a) ABB and ABC t-MBT lattice structures. The dotted lines between Mn atoms are drawn to indicate the different stacking.  The color coding of the atoms is the same as in Fig.\ref{fig:param_unit_cell}.  These figures were created with VESTA~\cite{vesta}. (b) Compatibility relations between the irreps of $D_{3d}$ and $C_{3v}$. }
		\label{fig:latt_compatibility}
	\end{center}
\end{figure}

For t-MBT, we now consider an ABB stacking pattern as depicted in Fig. \ref{fig:latt_compatibility}(a), which could be fabricated by placing individual SLs on top of each other. The first two layers preserve the stacking pattern of the bulk, but the top layer Mn atoms are placed directly on top of the Mns atoms of the middle layer. This structure belongs to the space group $P3m1$ ($156$), with point group is $C_{3v}$ at the $\Gamma$ point~\cite{isotropy, fyndsim}. $C_{3v}$ posses identity operation, three-fold rotational symmetry, and three mirror symmetries. This stacking pattern breaks inversion symmetry, opening the possibility to obtain a second harmonic generation signal in t-MBT, contrary to the case of d-MBT, where all the stacking patterns considered preserved inversion. The compatibility relations between the irreps of $D_{3d}$ and  $C_{3v}$~\cite{genpos} are shown in Fig. \ref{fig:latt_compatibility}. These relations are relevant to describe the symmetry breaking process, and also apply to the description of the effect on the symmetries of an applied static electric field in the $z$-direction, which also breaks inversion symmetry. The representation of the lattice vibration is $\Gamma_{vib.} = 21 A_1 \oplus 21 E$ and due to the absence of inversion symmetry both irreps in $\Gamma_{vib.}$ are Raman and infrared active. For the point group $C_{3v}$, the Raman tensors are shown in Eqs. \eqref{eq:raman_c3v1}-\eqref{eq:raman_c3v2}~\cite{loudon1964}. 
\begin{align} \label{eq:raman_c3v1}
R(A_{1}) & = \begin{pmatrix}
a & 0 & 0\\
0 & a & 0\\
0 & 0 & b
\end{pmatrix}\\
R^{(a)}(E) & = \begin{pmatrix}
0 & c & d\\
c  & 0 & 0\\
d & 0 & 0
\end{pmatrix}, R^{(b)}(E) & = \begin{pmatrix}
c & 0 & 0\\
0 & -c & d\\
0 &  d & 0
\end{pmatrix}.
\label{eq:raman_c3v2}
\end{align}

In summary, in this section we analyzed the group theory aspects of the lattice vibrations in few-layer MBT, and considered different deviations from the bulk stacking pattern. In the next section, we consider the group theory aspects of the magnetic transition below the critical temperature.

\section{Magnetic order in \lowercase{d}-MBT}
\label{sec:magneticmbt}

In this section, we consider the symmetry aspects of the magnetic transition in MBT, focusing on AB stacked d-MBT because this configuration already exhibits topological order linked with the magnetic order~\cite{Li_2019}.  First, we discuss the implications of the lattice on the symmetry-allowed magnetic orders. Then we derive a simple spin model for d-MBT and conclude this section discussing the possible coupling between magnetic order and phonons, which elucidates possible mechanisms to manipulate the magnetic order and therefore the topology of the system.

\subsection{Symmetry-allowed magnetic order}

As discussed in the previous sections, d-MBT has space group $P \bar 3 m 1$ $(164)$ with a $D_{3d}$ point group at the $\Gamma$ point. The Mn magnetic atoms are located at the $(2d)$ Wyckoff's positions with coordinates $(1/3,2/3,z)$ and $(2/3,1/3,-z)$. 

\begin{figure}[t]
	\begin{center}
		\includegraphics[width=6.5cm]{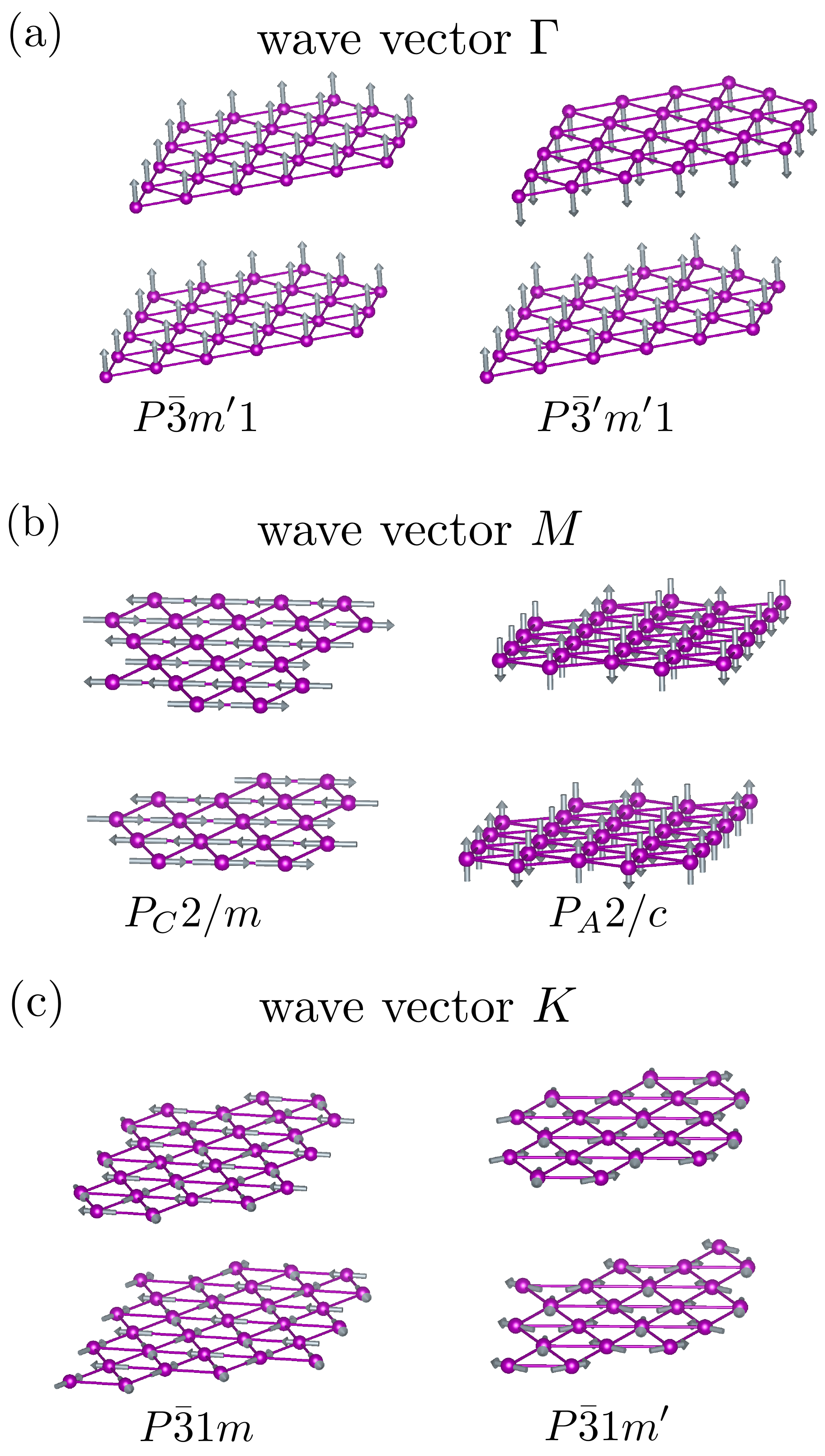}
		\caption{(Color online.) Symmetry-allowed magnetic moment configuration for three different high-symmetry wave vectors: (a)  $\boldsymbol k = (0,0,0)$ ($\Gamma$), (b) $\boldsymbol k = (1/2,0,0)$ ($M$), and (c) $\boldsymbol k = (1/3,1/3,0)$ ($K$). The corresponding magnetic space group is shown below.  The grey arrows indicate the magnetic moment orientations.}
		\label{fig:mag_figs}
	\end{center}
\end{figure}

The relevant magnetic space group is $P \bar 3 m 11'$, which is obtained from the crystallographic space group $P \bar 3 m 1$ by addition of its properties under a time-reversal operation $\mathcal T$. Different magnetic orders are allowed depending on the wave vector of the magnetic order relevant for the transition, which is determined by the magnetic unit cell considered. For example, if the chemical unit cell and the magnetic unit cell are identical, the associated wave vector is $\boldsymbol k = (0,0,0)$. In Appendix \ref{app:mag_tables} we show the subgroups of $P \bar 3 m 11'$ associated with three high-symmetry points in the Brillouin zone: $\Gamma$, $M$, and $K$. Each of the subgroups is associated with the onset of a particular magnetic configuration. In Fig. \ref{fig:mag_figs}, we present representative cases for these three wave vectors and plot the associated magnetic order. 

Next, we discuss in detail the magnetic transitions associated with the wave vector $\boldsymbol k = (0,0,0)$, which corresponds to cases where the chemical unit cell and the magnetic unit cell are identical. The irreducible correpresentations (IC) are shown in Table \ref{tab:IC}, and were obtained following Ref.~\onlinecite{toledano1987landau}. The magnetic order associated with each of the ICs listed in Table \ref{tab:IC} can be derived by inspecting the transformations of the magnetic moments under the operation of the group, recalling that it transforms as an axial vector localized on the Mn atoms. Alternatively, the magnetic order can be derived with \small{ISOTROPY}~\cite{isotropy}. The $A_{1g}$  IC corresponds to a paramagnetic phase.  The $A_{2g}$  and $E_{g}$ ICs are associated with out of plane and in-plane FM order respectively, while $A_{1u}$  and $E_{u}$ correspond to the AFM order. The $A_{2u}$ IC is not associated with magnetic order for the (2d) Wyckoff's position occupied by the Mn atoms.

\begin{table}[]
\begin{tabular}{|c|c|c|}
\hline Subgroup & Order & Index \\
\hline ${D}_{3 {d}}(-3 \mathrm{m})$ & 12 & 1 \\
\hline ${C}_{3 {v}}(3 \mathrm{m})$ & 6 & 2 \\
\hline ${D}_{3}(32)$ & 6 & 2 \\
\hline ${C}_{3 {i}} (-3)$ & 6 & 2 \\
\hline ${C}_{3}(3)$ & 3 & 4 \\
\hline ${C}_{2 {h}}(2 / {m})$ & 4 & 3 \\
\hline ${C}_{{s}}({m})$ & 2 & 6 \\
\hline ${C}_{2}(2)$ & 2 & 6 \\
\hline  ${C}_{{i}}(-1)$ & 2 & 6 \\
\hline ${C}_{1}(1)$ & 1 & 12 \\
\hline
\end{tabular}
\caption{Subgroups of $D_{3d}$. }
\label{tab:subsd3d}
\end{table}

The point group $D_{3d}$ possess three subgroups of index two (they contain half of the elements of the group), as shown in Table \ref{tab:subsd3d}: ${C}_{3 {v}}$, $D_3  = \{ \mathbb{1}, 3_z, 3_z^{-1}, 2_x, 2_{xy}, 2_y\}$, and ${C}_{3 {i}}$. These groups are relevant because they are used to construct the \textit{black and white} magnetic point groups $M$ (Shubnikov point groups of the third kind) to describe magnetic order. They have the group structure $M = H +  (D_{3d}-H)\mathcal T$, with $H={C}_{3 {v}}, {D}_{3}$, or ${C}_{3 {i}}$~\cite{toledano1987landau}. Additionally, we can construct the gray point group $M = D_{3d} +  D_{3d}\mathcal T$ associated with the paramagnetic phase, since $\mathcal T$ is a symmetry of the group.

\begin{table}[b]
\begin{tabular}{|c|c|c|c|}
\hline
Magnetic moments & $\{ C^+_{3z} | (0,0,0)\} $  & $\{ C_{2x} | (0,0,0)\} $ & $\{ \mathcal I | (0,0,0)\}$ \\ \hline
$(M_x,M_y,M_z)$  & $(-1,-1,1)$ & $(1, -1,-1)$      &   $(1,1,1)$     \\ \hline
$(L_x,L_y,L_z)$  & $(-1,-1,1)$ & $(-1,1,1)$         &   $(-1,-1,-1) $    \\ \hline
\end{tabular}
\caption{Transformation properties of the axial vectors $\boldsymbol M$ and $\boldsymbol L$ under the generators of the space group $P \bar 3 m 1$.}
\label{tab:trans}
\end{table}

Now, we derive the terms for a spin model to describe the transition from the paramagnetic to the magnetically ordered state in d-MBT. For this, we follow the discussion presented in Refs.~[\onlinecite{toledano1987landau,rado1963magnetism}], which is restricted to transitions where the chemical and magnetic unit cell are the same~\cite{rado1963magnetism}, valid for ordering wave vector $\boldsymbol k = (0,0,0)$.  As mentioned before, the two Mn atoms in the unit cell are located at the Wycoff's positions (2d) with coordinates $(1/3,2/3,z)$,  $(2/3,1/3,-z)$. We label the magnetic moments localized at each of the Mn atoms as $\boldsymbol s_1$ and $\boldsymbol s_2$, respectively, and define the two vectors
\begin{align}
\boldsymbol M&  = \boldsymbol s_1 +  \boldsymbol s_2, \\
\boldsymbol L &  = \boldsymbol s_1 -  \boldsymbol s_2, 
\end{align}
which represent the total magnetization and the AFM order parameter in the unit cell, respectively. First, we need to study the transformation properties of $\boldsymbol M$ and $\boldsymbol L$ under the operations of the group. For this, it is sufficient to consider the generators of $P \bar 3 m 1$, which are~\cite{gtpack1,gtpack2} 
$\{ \mathcal I | (0,0,0)\}$,
$\{ C^+_{3z} | (0,0,0)\} $,
$\{ C_{2x} | (0,0,0)\} $, and
$\{ \mathcal I | (0,0,0)\}$.
Consider first the two-fold rotation $\{ C_{2x} | (0,0,0)\}$, which switches the position of $\boldsymbol s_1$ and $\boldsymbol s_2$. The spin components $s^z_{1,2}$ and $s^y_{1,2}$ switch sign, while leaving $s^x_{1,2}$ unchanged. With this transformation properties, we find that  $ \{ C_{2x} | (0,0,0)\}  \boldsymbol M = (M_x, -M_y, -M_z)$  and 
$ \{ C_{2x} | (0,0,0)\}  \boldsymbol L = (-L_x, L_y, L_z)$. On the other hand, the three-fold rotation 
$\{ C^+_{3z} | (0,0,0)\} $ does not interchange $\boldsymbol s_1$ and $\boldsymbol s_2$. The spin components $s^z_{1,2}$  are parallel to the three-fold rotation axis, and do not change sign. Meanwhile, both $s^x_{1,2}$ and $s^y_{1,2}$ components change sign. With this transformation properties, we find $\{ C^+_{3z} | (0,0,0)\} \boldsymbol M = (-M_x, -M_y, M_z)$ and $\{ C^+_{3z} | (0,0,0)\} \boldsymbol L = (-L_x, -L_y, L_z)$. Inversion $\{ \mathcal I | (0,0,0)\}$ sends an atom on the top layer to the bottom layer, $\boldsymbol s_1 \leftrightarrow \boldsymbol s_2$ and does not affect the axial vectors. 

In Table \ref{tab:trans}, we summarize the transformation properties of each of the components of $\boldsymbol M$ and $\boldsymbol L$. Interactions are allowed only among components that transform in the same way under the operations of the group. From the results in Table \ref{tab:trans}, the following quadratic terms are totally symmetric: $M^2_x$, $M^2_y$, $M^2_z$, $L^2_x$, $L^2_y$ and $L^2_z$. We find that no cross terms $M_i L_j$, $M_i M_j$, or $Li Lj$ are allowed by symmetry. These terms are associated with relativistic effects and can lead to canted states~\cite{toledano1987landau}. With these results, we can construct the Landau free energy up to fourth order, $F_{\text{mag}} = F_0 + \alpha_M/2 M^2+\beta_M/4 M^4+\alpha_L/2 L^2+\beta_L/4 L^4$, where $F_0$ represents the equilibrium free energy, $M = |\boldsymbol M|$, $L = |\boldsymbol L|$ and the coefficients $\alpha_i, \beta_i$ with $i=M,L$ are real numbers.Therefore, from these considerations, we expect either pure FM or pure AFM (collinear) order, depending on the details of the energetics.

\subsection{Coupling between the magnetic order and phonons}

In the previous subsection, we determined a simple model to describe the magnetic order in d-MBT. Here, we study the symmetry-allowed coupling between the phonons and the magnetic order, which suggests possible ways to manipulate the magnetic order with phonons driven by strong laser pulses. As discussed in the introduction, such mechanisms have been used to manipulate magnetic and other correlated states such as superconductivity and ferroelectricity~\cite{forst2011,FORST201324,Fausti189,mitrano2016,Nova2017,fechner2016,Nova1075,Mankowsky2014,mciver2018lightinduced,mankowsky2015}. In particular, in Ref.~\onlinecite{Nova2017}, a protocol involving driving two IR modes with strong laser pulses out of phase was employed to simulate an applied magnetic field and excite spin precession in ErFeO$_3$.

We consider the scenario where an IR mode with symmetry $A_{2u}$ is excited with a strong laser pulse. The strongly driven IR mode will couple non-linearly to the Raman modes. In particular, we consider coupling to an $A_g$ modes, for simplicity. We will assume d-MBT presents AFM order with representation $A_{1u}$, such that the system is described by the Ginzburg-Landau (GL) model derived with \small{ISOTROPY}
\begin{equation}
\mathcal F(t) =\mathcal F_0 + \mathcal F_{\text{mag}} + \mathcal F_{\text{ph}} + \mathcal F_{\text{mag-ph}}+\mathcal F_{\text{L}}(t),
\end{equation}
where $\mathcal F_0$ is the equilibrium GL term, the magnetic contribution is taken to be $ \mathcal F_{\text{mag}} = \alpha_L/2 L^2+\beta_L/4 L^4 $, the phonon sector is given by $\mathcal F_{\text{ph}} = \Omega_{IR}^2/2 Q_{IR}^2 + \Omega_{R}^2/2 Q_{R}^2 + a/3 Q_{R}^3 + b Q_{IR}^2 Q_{R} + \text{(quartic terms)}$ where $Q_{IR}$ and $Q_{R}$ correspond to the IR and Raman phonons with frequencies $\Omega_{IR}$ and $\Omega_{R}$, respectively. The coupling with the magnetic order is dictated by $\mathcal F_{\text{mag-ph}} = L^2 ( c Q_{R}+d  Q^2_{R}+e Q^2_{IR})$. Finally, $\mathcal F_{\text{L}} = F^{i} \Phi(t) \sin \left(\Omega t\right) Q_{IR}$ describes the driving term where $F$ characterizes the amplitude of the laser, and the time-dependent part defines the laser pulse profile at frequency $\Omega$  with Gaussian shape $\Phi(t)=e^{-t^2/(2 \sigma^2)}$. All the coefficients are assumed to be real, In particular, the numerical values for the coefficients require the implementation of first-principles calculations. This model is presented with the main goal of highlighting that symmetry allows the lattice vibrations to couple with the magnetic order. Furthermore, non-linear phononics processes induced by strong laser process provide a mechanism to possibly manipulate indirectly the magnetic order and thus the topology.

In closing, is important to mention that recent magnetic force microscopy measurements have revealed the presence of AFM domain walls at the surface of MBT~\cite{Sass2020}, which are not captured in the present analysis. Furthermore, other effects such as single-ion anisotropies, the finite size of the sample, and surface interactions are not taken into consideration.  These are natural topics for future studies.

\section{Conclusions}
\label{sec:conclusions}

In conclusion, we presented a group theory study of the vibrational modes in bulk MnBi$_2$Te$_4$ (MBT) and few septuple-layer MBT. Also, we presented a detailed group theory analysis of the magnetic order in d-MBT, which makes a connection with possible mechanisms to manipulate the magnetic order aided by the phonons. Our main findings for paramagnetic bulk MBT are: i) we found the phonon selection rules, ii) the expected phonon degeneracies, and iii) the real-space lattice displacements that bring the dynamical matrix into a block-diagonal form. 

For few-layers MBT, our main findings are: i) the lattice vibration representation for systems with up to five layers, summarized in Table \ref{tab:irreps_N}, which indicate the number of phonons expect for each system and their degeneracy; ii) the phonon selection rules for all the few-layer systems considered. For single layer MBT, we considered the effect of strain, possibly induced by a substrate, and discussed its effect on the vibrational modes. Also, we found the compatibility relations for the phonon modes. 

For double layer MBT, we obtained: i) the space group for structures that present a stacking different from the ABC bulk stacking, relevant when constructing stacks of MBT out of single layers; ii) discussed the symmetries of the resulting lattice in each case, iii) obtained the representation of the vibrational modes; iv) and showed that the real-space displacements allow for motion of the Mn atoms in Raman modes, as opposed to bulk. The last point is relevant for the possible manipulation of the magnetic order, since motion between Mn atoms can more easily modify exchange pathways and impact the order. 

For trilayer MBT, we additionally considered an ABB stacking which presents broken inversion symmetry, which will exhibit a non-zero second-harmonic generation signal and can be used to characterize this stacking pattern. Our results can be used to help interpret Raman measurements on few-layer MBT, and to guide the construction of van der Waals heterostructures with specific desired symmetries. 

We also discussed the group theory aspects of the magnetic order in double layer MBT as a representative case where the magnetic order and topology are already linked. We obtained a model to describe the magnetic transition taking into account the symmetries of the lattice. Furthermore, we showed that the symmetries of the lattice allow for coupling between the magnetic order and the lattice vibrations, opening up the possibility to use non-linear phononic processes to tune the magnetic order and topology.

The group theory analysis of the magnetic order and vibrational modes we presented here should lend themselves to being used as the basis of a more refined analysis taking into account surface effects. Such an extension should help understand the relationship between the band topology and magnetic order in MBT, and other symmetry-related materials.

\section{Acknowledgements} 
We thank A. Ernst for useful discussions. This research was primarily supported by the National Science Foundation through the Center for Dynamics and Control of Materials: an NSF MRSEC under Cooperative Agreement No. DMR-1720595, with additional support from NSF DMR-1949701. A.L. acknowledges support from the funding grant: PID2019-105488GB-I00. 

%

\appendix

\section{Lattice vibrational modes for d-MBT}
\label{app:biSL}

In this appendix, we show the real-space displacements for the two-dimensional representations $E_g$ and $E_u$ of d-MBT with AB stacking.

\begin{figure}[H]
	\begin{center}
		\includegraphics[width=8.5cm]{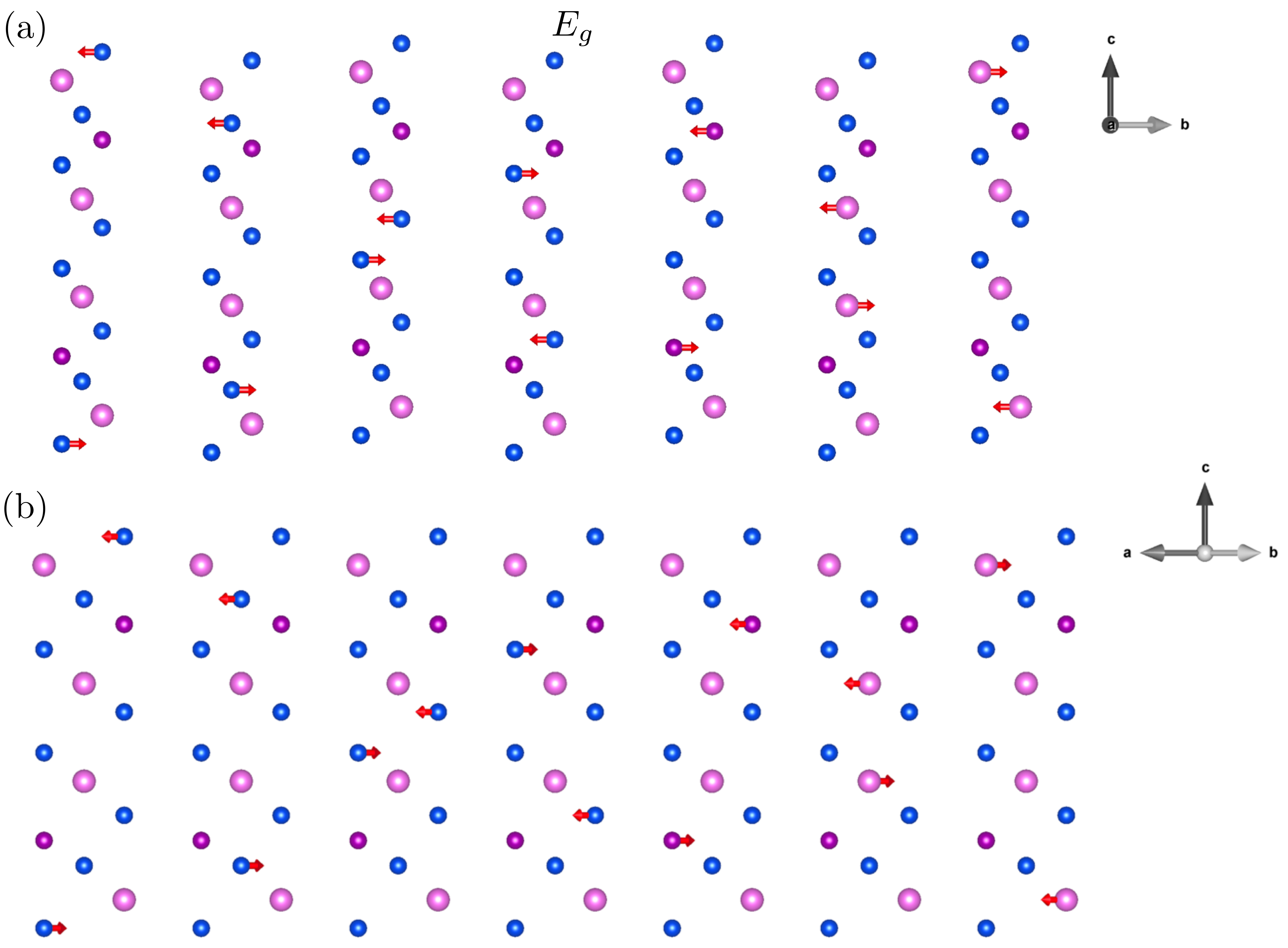}
		\caption{(Color online) Lattice vibrational modes with irreps $E_{g}$ for d-SL MBT in the paramagnetic phase. The color coding of the atoms is the same as in Fig.\ref{fig:param_unit_cell}.  This figure was created with VESTA~\cite{vesta}. }
		\label{fig:latt_vib_bi}
	\end{center}
\end{figure}

\begin{figure}[H]
	\begin{center}
		\includegraphics[width=8.5cm]{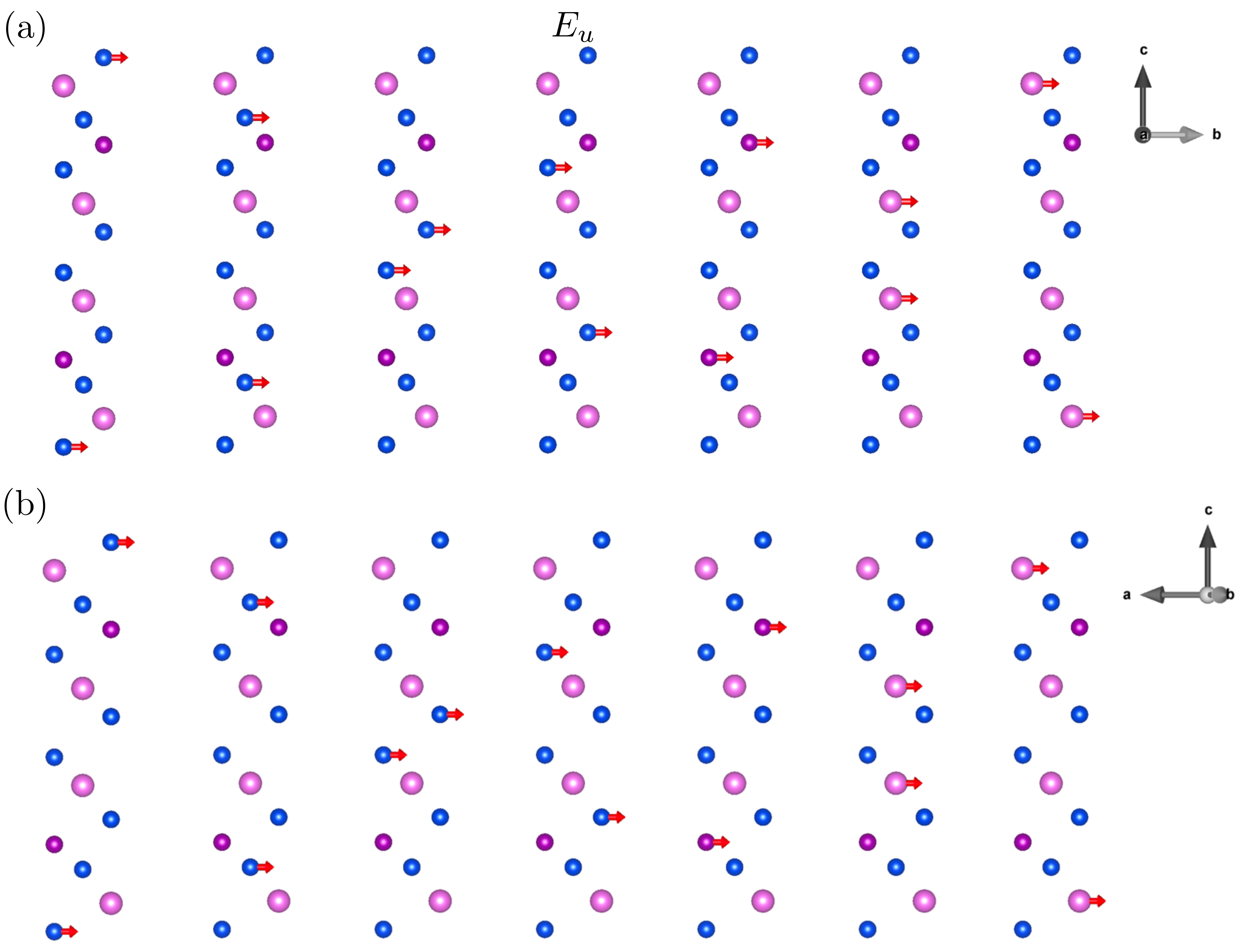}
		\caption{(Color online) Lattice vibrational modes with irreps $E_{u}$ for d-SL MBT in the paramagnetic phase.  The color coding of the atoms is the same as in Fig.\ref{fig:param_unit_cell}. This figure was created with VESTA~\cite{vesta}. }
		\label{fig:latt_vib_bi}
	\end{center}
\end{figure}

\section{Character tables}
\label{app:tables}

In this appendix, we list the character tables for point groups considered in this work.

\begin{table}[H]
\centering
\begin{tabular}{|c|c|c|c|c|c|c|c|}
\hline
$D_{3d}$ & $E$ & $2C_3$ & $3C'_{2}$ & $i$ & $2S_6$ & $3\sigma_d$ & functions                  \\ \hline
$A_{1g}$ & 1   & 1      & 1        & 1   & 1      & 1           & $x^2+y^2, z^2$             \\ \hline
$A_{2g}$ & 1   & 1      & -1       & 1   & 1      & -1          &                            \\ \hline
$E_g$    & 2   & -1     & 0        & 2   & -1     & 0           & $x^2-y^2$, $xy$, $xz$,$yz$ \\ \hline
$A_{1u}$ & 1   & 1      & 1        & -1  & -1     & -1          &                            \\ \hline
$A_{2u}$ & 1   & 1      & -1       & -1  & -1     & 1           & $z$                        \\ \hline
$E_u$    & 2   & -1     & 0        & -2  & 1      & 0           & $x,y$                      \\ \hline
\end{tabular}
\caption{$D_{3d}$ point group character table. }
\label{tab:d3d}
\end{table}

\begin{table}[H]
\centering
\begin{tabular}{|c|c|c|c|c|c|}
\hline & ${E}$ & ${C}_{{2}}$ & ${i}$ & ${\sigma}_{{h}}$ & functions \\
\hline ${A}_{{g}}$ & 1 & 1 & 1 & 1 & ${R}_{{z}}$ , ${x}^{2}, {y}^{2}, {z}^{2}, {x y}$ \\
\hline ${B}_{{g}}$ & 1 & -1 & 1 & -1 & ${R}_{{x}}, {R}_{{y}}$ , ${x z}, {y z}$ \\
\hline ${A}_{{u}}$ & 1 & 1 & -1 & -1 & ${z}$  \\
\hline ${B}_{{u}}$ & 1 & -1 & -1 & 1 & ${x}, {y}$  \\
\hline
\end{tabular}
\caption{$C_{2h}$ point group character table. }
\label{tab:d3d}
\end{table}

\begin{table}[H]
\centering
\begin{tabular}{|c|c|c|c|c|}
\hline & ${E}$ & ${2 C}_{{3}}$ & ${3} {\sigma}_{{v}}$ & functions \\
\hline ${A}_{{1}}$ & 1 & 1 & 1 & ${z}$, ${x}^{2}+{y}^{2}, {z}^{2}$ \\
\hline ${A}_{{2}}$ & 1 & 1 & -1 & ${R}_{{z}}$  \\
\hline ${E}$ & 2 & -1 & 0 & $({x}, {y})\left({R}_{{x}}, {R}_{{y}}\right)$,$\left({x}^{2}-{y}^{2}, {x y}\right)({x z}, {y z})$ \\
\hline
\end{tabular}
\caption{$C_{3v}$ point group character table. }
\label{tab:d3d}
\end{table}

\section{Subgroups of $P \bar 3 m 1$ at different wave vectors.}
\label{app:mag_tables}

In this appendix we list the subgroups obtained as a result of induced magnetic order with the high-symmetry wave vectors $\boldsymbol k=(0,0,0)$ ($\Gamma$ point), $\boldsymbol k=(1/2,0,0)$ (M point), and $\boldsymbol k=(1/3,1/3,0)$ (K point).

\begin{table}[H]
\centering
\begin{tabular}{|c|c|c|}
\hline
Irrep          & Subgroup                                              &    Order Parameter Direction      \\ \hline
$A_{1g}$   &  $P \bar 3 m 1  $ (164.85)                   &  P1  (a)  \\ 
$A_{2g} $  &  $P \bar 3m' 1  $ (164,89)                     &  P1  (a)  \\ 
$E_g$       &   $C2/m $    (12.58)                               &  P1   (a, -1.732a )\\ 
$ E_g $     &  $C 2^{\prime} / {m}^{\prime} $ (12.62) &  P 2 (a, 0.577 a)\\ 
$ E_g $     &  $P \bar 1 $  (2.4)                                  &  C1 (a, b) \\ 
$ A_{1u} $ &  $P \bar 3' m' 1  $  (164.88)                   &  P1  (a) \\ 
$ A_{2u} $ &  $ P \bar 3 ' m1 $ (164.87 )                   &  P1  (a) \\ 
$ E_{u} $   &  $C 2^{1} / {m} $ (12.60)                       &  P1 (a,-1.732 a)\\ 
$ E_{u} $   &  $  C2/m' $(12.61)                                   & P2  (a, 0.577a) \\ 
$ E_{u} $   &  $  P \bar 1^{\prime} $ (2.6)                  &  C1  (a, b) \\ 
\hline
\end{tabular}
\caption{Subgroups of $P \bar 3 m 1$ at wave vector $\boldsymbol k=(0,0,0)$ obtained with \small{ISOTROPY}. The columns indicates: the irreps, the magnetic space group, and the order parameter direction in representation space. Each irrep leads to a particular magnetic moment ordering. We did not constraint the magnetic moments at the magnetic atoms Wyckoff's positions.}
\label{tab:Mpoint}
\end{table}

\begin{table}[H]
\centering
\begin{tabular}{|c|c|c|}
\hline
Irrep  & Subgroup          &    Order Parameter Direction      \\ \hline
$M_1^+$  & $P_C2/m$ (10.49 )   &P1  (a,0,0) \\
$M_1^+ $ &  $C_a2/m$ (12.64)  &P2  (a,a,0)                                 \\
$M_1^+$  &  $P\bar3m14$ (164.85)  &P3  (a,a,a) \\
$M_1^+$ & $P_S\bar1 $ ( 2.7)        & C1  (a,b,0) \\
$M_1^+ $ & $C2/m$ (12.58 )      & C2  (a,a,b) \\
$M_1^+ $ & $ P\bar1$  (2.4)              &   S1  (a,b,c)\\ 
$M_2^+$  &  $P_A2_1/c$ (14.83) & P1  (a,0,0) \\
$M_2^+ $  &  $C_a2/m$ (12.64)    &P2  (a,a,0) \\
$M_2^+ $ & $P\bar3m'1$ (164.89 )  &P3  (a,a,a) \\
$M_2^+ $  &   $P_S\bar1$ (2.7)     &C1  (a,b,0) \\
$M_2^+$   &  $C2'/m' $ (12.62) &C2  (a,a,b) \\
$M_2^+$   &  $P\bar1$ (2.4)       &S1  (a,b,c) \\
$M_1^-$   &  $P_A2/c $ (13.73) &P1  (a,0,0)      \\ 
$M_1^-$  &  $C_a2/m $(12.64)  &P2  (a,a,0) \\
$M_1^-$  &$P\bar3'm'1$ (164.88 ) &P3  (a,a,a) \\
$M_1^- $ &  $P_S\bar1 $ (2.7)    &C1  (a,b,0) \\
$M_1^-$  & $C2/m' $ (12.61) & C2  (a,a,b) \\
$M_1^-$  &  $P\bar1'$ (2.6)       &S1  (a,b,c) \\
$M_2^-$  &  $P_C2_1/m$(11.57) &P1  (a,0,0)       \\ 
$M_2^-$   &$C_a2/m$ (12.64 )   &P2  (a,a,0) \\
$M_2^-$  & $P\bar3'm1$ (164.87)  &P3  (a,a,a) \\
$M_2^-$  & $P_S\bar1  $  (2.7 )   &C1  (a,b,0) \\
$M_2^-$  & $C2'/m$ (12.60 ) &C2  (a,a,b) \\
$M_2^-$ &   $P\bar1'$ (2.6)           &S1  (a,b,c) \\
\hline
\end{tabular}
\caption{Subgroups of $P \bar 3 m 1$ at wave vector $\boldsymbol k=(1/2,0,0)$ obtained with \small{ISOTROPY}. The columns indicates: the irreps, the magnetic space group, and the order parameter direction in representation space. Each irrep leads to a particular magnetic moment ordering. We did not constraint the magnetic moments at the magnetic atoms Wyckoff's positions.}
\label{tab:Mpoint}
\end{table}

\begin{table}[H]
\centering
\begin{tabular}{|c|c|c|}
\hline
Irrep  & Subgroup          &    Order Parameter Direction      \\ \hline
$ K_1$ &  ${P}\bar 31 \mathrm{m}$  (162.73)                        & P1  (a, 0) \\ 
$ K_1$ &  ${P} \bar 3^{\prime} 1 \mathrm{m}^{\prime}$ (162.76 ) & P2  (0, a) \\ 
$ K_1$ &  ${P} 312 $ 149.21   & C1  (a,b) \\ 
$ K_2$ &  $ {P}\bar3^{\prime} 1 \mathrm{m} $ (162.75) &  P1  (a, 0) \\ 
$ K_2$ &  $P \bar 31 \mathrm{m}^{\prime} $ (162.77) &P2  (0,a) \\ 
$ K_2$ &  $ P 312^{\prime} (149.23) $ & C1 (a,b) \\ 
$ K_3$ & $ P 31 \mathrm{m} $ (157.53) & P 1  (a, 1.732 a,-1.732 a, a) \\ 
$ K_3$ &  ${P} 31 \mathrm{m}^{\prime}$ (157.55) & $\mathrm{P} 2  (\mathrm{a},-0.577 \mathrm{a}, 0.577 \mathrm{a}, \mathrm{a}) $\\ 
$ K_3$ &  ${C} 2 / \mathrm{m} $(12.58) & $\mathrm{P} 4  (\mathrm{a}, 1.732 \mathrm{a}, 0,0)$ \\ 
$ K_3$ & $ {C} 2^{\prime} / \mathrm{m}$ (12.60) & $ \mathrm{P} 7  (0,0, \mathrm{a},-0.577 \mathrm{a})$ \\
$  K_3$ &  ${C} 2^{\prime} / \mathrm{m}^{\prime}   $ (12.62) &$\mathrm{P} 8 (\mathrm{a},-0.577 \mathrm{a}, 0,0) $\\ 
$ K_3$ &  ${C} 2 / \mathrm{m}^{\prime} $ (12.61) & $\mathrm{P} 9  (0,0, \mathrm{a}, 1.732 \mathrm{a})$ \\ 
$ K_3$ & $ {P} 3 $ (143.1) & $\mathrm{C} 1  (\mathrm{a}, \mathrm{b},-\mathrm{b}, \mathrm{a})$ \\ 
$ K_3$ &${Cm} ( 8.32 ) $&$ \mathrm{C} 2  (\mathrm{a}, 1.732 \mathrm{a}, \mathrm{b},-0.577 \mathrm{b}) $\\ 
$ K_3$ &  ${Cm}^{\prime} (8.34) $& $\mathrm{C} 4  (\mathrm{a},-0.577 \mathrm{a}, \mathrm{b}, 1.732 \mathrm{b}) $\\  
$ K_3$ &  ${P}\bar 1 $ (2.4) & $\mathrm{C} 10  (\mathrm{a}, \mathrm{b}, 0,0) $\\ 
$ K_3$ &  ${C} 2$ (5.13) & $\mathrm{C} 11  (\mathrm{a}, 1.732 \mathrm{a}, \mathrm{b}, 1.732 \mathrm{b}) $\\ 
$ K_3$ &  ${C} 2^{\prime} (5.15) $&$ \mathrm{C} 15  (\mathrm{a},-0.577 \mathrm{a}, \mathrm{b},-0.577 \mathrm{b})$ \\  
$ K_3$ &  ${P}\bar1^{\prime} $ (2.6) &$ \mathrm{C} 16  (0,0, \mathrm{a}, \mathrm{b})$ \\ 
$ K_3$ &  ${P} 1$  (1.1)  &  $ 4\mathrm{D} 1 (\mathrm{a}, \mathrm{b}, \mathrm{c}, \mathrm{d})$\\
\hline
\end{tabular}
\caption{Subgroups of $P \bar 3 m 1$ at wave vector $\boldsymbol k=(1/3,1/3,0)$ obtained with \small{ISOTROPY}. The columns indicates: the irreps, the magnetic space group, and the order parameter direction in representation space. Each irrep leads to a particular magnetic moment ordering. We did not constraint the magnetic moments at the magnetic atoms Wyckoff's positions.}
\label{tab:Mpoint}
\end{table}

\end{document}